\documentclass[10pt,a4paper,twocolumn,english]{article}
\usepackage[T1]{fontenc}
\usepackage[latin9]{inputenc}
\usepackage{array}
\usepackage{units}
\usepackage{textcomp}
\usepackage{multirow}
\usepackage{amstext}
\usepackage{graphicx}
\usepackage{setspace}

\makeatletter

\pdfpageheight\paperheight
\pdfpagewidth\paperwidth

\providecommand{\tabularnewline}{\\}

 \usepackage[small,bf]{caption}

\makeatother

\usepackage{babel}
\begin{document}

\title{Statistical anomalies in 2011--2012 Russian elections\\
 revealed by 2D correlation analysis}

\author{Dmitry Kobak\\
Imperial College London, UK\\
\\
Sergey Shpilkin\\
\\
Maxim S. Pshenichnikov}

\maketitle
\textbf{Here we perform a statistical analysis of the official data
from recent Russian parliamentary and presidential elections (held
on December 4th, 2011 and March 4th, 2012, respectively). A number
of anomalies are identified that persistently skew the results in
favour of the pro-government party, United Russia (UR), and its leader
Vladimir Putin. The main irregularities are: (i) remarkably high correlation
between turnout and voting results; (ii) a large number of polling
stations where the UR/Putin results are given by a round number of
percent; (iii) constituencies showing improbably low or (iv) anomalously
high dispersion of results across polling stations; (v) substantial
difference between results at paper-based and electronic polling stations.
These anomalies, albeit less prominent in the presidential elections,
hardly conform to the assumptions of fair and free voting. The approaches
proposed here can be readily extended to quantify fingerprints of
electoral fraud in any other problematic elections.}

\bigskip{}

Legislative elections to the Russian Parliament, the Duma, and presidential
elections were held in Russia on December 4th, 2011 and March 4th,
2012, respectively. Widespread belief that the outcome of legislative
elections was manipulated led to large-scale public protests unseen
in Russia since the early 90s; still, virtually none of the alleged
manipulations were officially acknowledged. Statistics is known to
be a powerful tool to pinpoint irregularities in election data that
could be caused by unfair or fraudulent voting \cite{Mikhailov2004,Myagkov2009,Klimek2012},
and this pair of major elections provides a unique opportunity for
comparing election data side-by-side, as most of the party leaders
later ran for president. On one hand, sociogeographic distribution
of the voters could not have substantially changed within three months
between the elections, so both datasets should exhibit similar patterns.
On the other hand, public protests after parliamentary elections resulted
in unprecedented anti-forgery activities at the presidential elections,
such as live web broadcast from most of the polling stations and intense
public control by volunteer observers. With this in mind, we, inspired
by methods of two-dimensional correlation spectroscopy \cite{Hamm2011},
analyse the data from both elections in Russia and identify a number
of anomalies that persistently skew the results in favour of the pro-government
party, United Russia (UR), and its leader Vladimir Putin.

\begin{figure*}[t]
\begin{centering}
\includegraphics[width=1\textwidth]{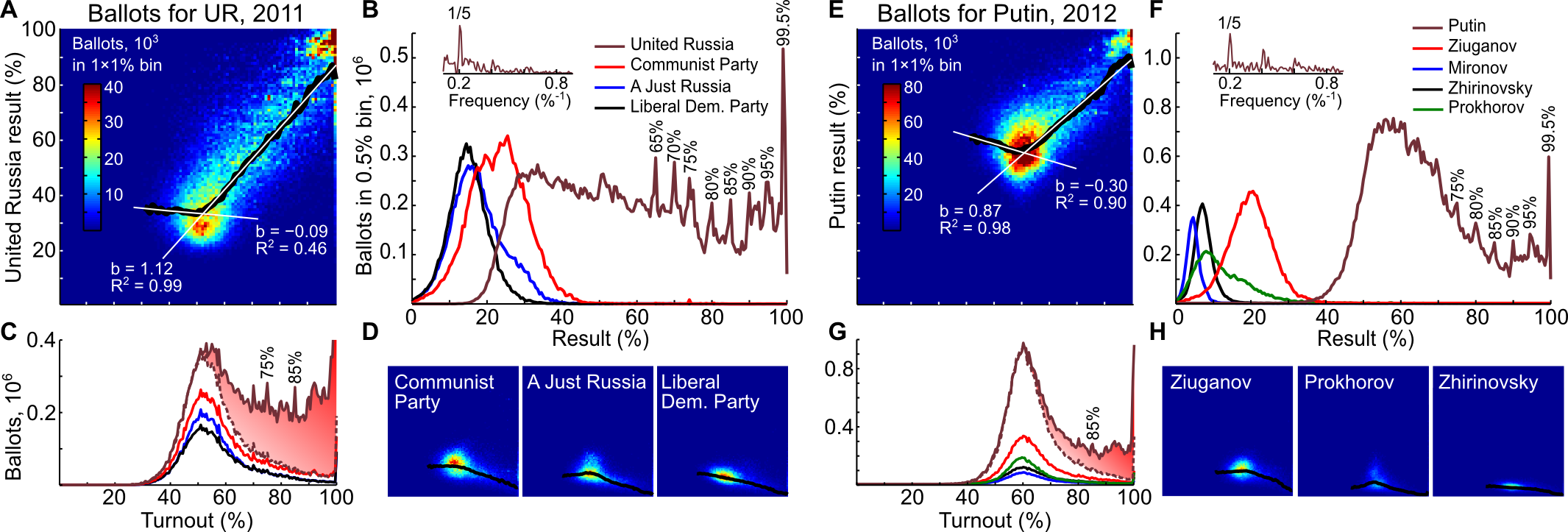}
\par\end{centering}

\caption{Summary of results by United Russia and Vladimir Putin. \textbf{(A)}
Ballots obtained at polling stations showing a certain turnout and
result of United Russia (in $1\%\times1\%$ bins). Number of ballots
is colour-coded; the cluster in the upper right corner is heavily
saturated to enable other data to be visible. The black curve depicts
an overall result for each turnout bin. White lines show linear fits
to the black curve before and after the 50\% turnout; the $R^{2}$
value and the regression coefficient are depicted next to each fit.
\textbf{(B)} Total number of ballots cast for each party depending
on the result at the polling station (in 0.5\% bins). Inset shows
the Fourier power spectrum of the United Russia trace. \textbf{(C)}
Number of ballots depending on the turnout (0.5\% bins). The colour
coding is the same as in (B). Dashed line shows the part of UR trace
proportional to the sum of all other parties; red shading shows the
difference. The UR trace is truncated at 100\% turnout for the sake
of clarity; the maximal value is $0.98\cdot10^{6}$. \textbf{(D)}
Two-dimensional histograms for three other elected parties. Colour
scale is the same as in (A). \textbf{(E--H)} Similar plots for the
presidential elections.}
\end{figure*}

The election data are officially available online at Russian Central
Election Committee website \emph{(izbirkom.ru)} detailed to a single
polling station. Seven parties participated in the parliamentary elections
with four of them having passed the 7\% threshold; five candidates
ran for president (see Methods for details). There are $\sim$95000
polling stations in Russia, grouped in 2744 constituencies in 83 regions.
The election statistics comprises more than 109 million of registered
voters with 65.7 and 71.7 million votes cast in legislative and presidential
elections, respectively. United Russia (UR) won the parliamentary
elections with a result of 49.31\%, while Vladimir Putin defeated
his rivals with a landsliding figure of 63.60\%.

\begin{spacing}{0.94999999999999996}
Figures 1A and 1E show 2D histograms of the number of ballots in favour
of UR/Putin as a function of turnout and respective vote share at
each polling station. Apart from the main clusters at $\sim$52\%
turnout and $\sim$30\% votes for UR and $\sim$60\%/55\% for Putin,
there are two prominent features at both plots that clearly distinguish
them from other participants\textquoteright{} histograms (Figs. 1D,H):
(i) an unusual cluster of votes in the vicinity of 95\% turnout, and
(ii) a long tail of votes beginning at the central peak which shows
a high correlation of the results with the turnout (marked by black
curves, known in 2D spectroscopy as the centre line slope \cite{Roy2011}).
The clusters at 90--100\% turnout yield $\sim$3.5 million ballots
for the winners in both elections and can be traced back to six republics
of North Caucasian Federal District, and Republics of Mordovia, Bashkortostan,
and Tatarstan. In each of these nine regions, there are a number of
constituencies that exhibit voting results with extremely low dispersion
across polling stations, significantly lower than dispersion value
imposed by binomial model (e.g., 25 constituencies with $p<0.0001$
for parliamentary and 9 constituencies for presidential elections,
see Table S1 and Methods). This suggests that the results in these
constituencies were artificially fixed to certain percentage values.
\end{spacing}

It is instructive to consider a projection of the 2D histograms onto
the vertical axis, which gives a distribution of the number of ballots
cast for UR and Putin depending on their results at every polling
station (Figs. 1B,F). The unique feature of these histograms is sharp
peaks located at {}``round'' numbers of 65\%, 70\%, 75\% etc. The
periodic character of these peaks is evident from the Fourier spectra
that show prominent harmonics at $\nicefrac{1}{5}\%^{-1}$ (insets).
By far the highest peak in both cases is located at 99.5\% and originates
solely from a single region of Chechen Republic. Other peaks can also
be traced back to particular constituencies, but are usually not confined
to a single region. These peaks, which are highly statistically significant
(see Table 1 and Methods), comprise $\sim$1.4 million ballots for
UR and $\sim$1.3 million ballots for Putin. The supernatural character
of the peaks strongly suggests that the votes for the winners were
manipulated a posteriori to fix the vote shares at appealing round
values.

The second prominent feature of the 2D histogram in Figs. 1A,E is
a remarkable correlation between the turnout and the result of UR
(correlation coefficient of 0.68) and Putin (0.53). Note that at lower
turnouts both correlations are negative, becoming positive only at
turnouts higher than the position of the main clusters. The histograms
for other competitors show exactly opposite behaviour: low or even
positive correlation at lower turnouts and negative correlation further
on (Figs. 1D,H). In general, correlation between turnout and voting
results is a well-known phenomenon, observed in many countries \cite{Hansford2010}.
However, dependencies as strong as found here are hard to explain
without an assumption of administrative pressure and/or vote manipulation
\cite{Klimek2012,Mikhailov2004,Myagkov2009}. 

The correlation between turnout and voting results at the national
scale could have arisen due to aggregation of widely dispersed but
otherwise uncorrelated results from different territories, given large
cultural and socio-economic differences between regions of Russia
as well as between urban and rural areas. To address this issue, the
data presented on Figs. 1A,E were decomposed into three parts: urban
areas, rural areas, and the nine aforementioned republics (see Fig.
S1 and Methods). Both urban and rural areas separately exhibit high
correlations; further detalization to the region level shows that
high correlation is not characteristic for every region but is confined
to only some regions of Russia. Furthermore, in regions demonstrating
high correlations, similar correlations are already observed at the
level of individual constituencies (see Supplementary Information).
This shows that the observed correlations are not an aggregation artifact
but an internal feature of specific constituencies (see SI).

\begin{figure}[t]
\begin{centering}
\includegraphics[width=1\columnwidth]{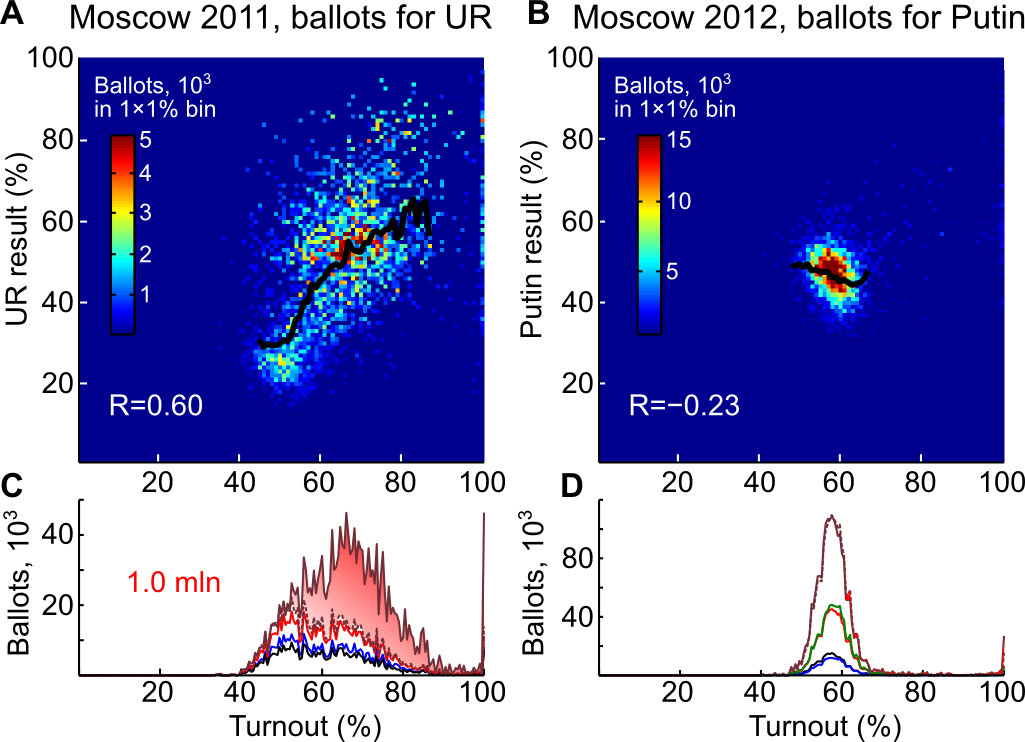}
\par\end{centering}

\caption{Voting results in the city of Moscow. Number of votes for UR \textbf{(A)}
and Putin \textbf{(B)} at polling stations showing a certain turnout
and result in parliamentary and presidential elections, respectively
(in $1\%\times1\%$ bins). R stands for correlation coefficient (excluding
5\% of ballots at highest and 5\% at lowest turnouts). Note that two
distinct clusters of ballots at $\sim$50\% and at $\sim$70\% turnout
and a high positive correlation between turnout and UR result in (A)
turned into a single well-confined cluster and negative correlation
between turnout and Putin\textquoteright{}s result in (B). \textbf{(C,
D)} Horizontal projections of (A) and (B), together with the histograms
of other participants. The colours are the same as in Fig.1. The red-coloured
number in (C) shows the area of the red shading, similar to Fig. 1C.}
\end{figure}

One of the most striking examples of such correlations is given by
the city of Moscow where parliamentary elections resulted in an extremely
high correlation between turnout and UR result (Fig. 2A). The situation
was totally reversed in the presidential elections, where Putin's
result was strongly anticorrelated with the turnout (Fig. 2B). Also,
the horizontal projections of the 2D histograms (which show the number
of ballots as a function of turnout) acquired similar shapes for all
candidates (Fig. 2D), in contrast to the parliamentary elections where
the UR curve had a pronounced tail at high turnouts (Fig. 2C). Moreover,
averaged standard deviation (SD) of the UR/Putin results across polling
stations in each Moscow constituency decreased sharply from 12\textpm{}5\%
(parliamentary elections, mean\textpm{}s.d.) to 4\textpm{}2\% (presidential
elections). This drastic change in the electoral data is most naturally
explained by the tight public control implemented by angry citizens
in Moscow after alleged falsifications in the parliamentary elections.

Moscow results demonstrate that dispersion across polling stations
in each constituency can serve as yet another metrics of election
anomalies. In urban constituencies one expects to find a relatively
uniform voting (i.e. with low dispersion) due to population homogeneity.
In both elections, there is a dense cluster of urban constituencies
(Fig. 3) showing SDs of around 2--7\%, which probably indicates the
normal range of SDs. At the same time, in the parliamentary elections
(Fig. 3A) there are many constituencies showing much larger SDs, up
to 27\% (see Table S2). Furthermore, there is a strong correlation
between the SD and the overall UR result (correlation coefficient
0.62), indicating that high SDs might be induced by manipulated results
at some (but not all) polling stations in a constituency. In contrast,
the similar data for the presidential elections (Fig. 3B) are much
more confined, with the number of constituencies with SD over 10\%
dropping from 185 to 28. Again, the most parsimonious explanation
is that in the presidential elections votes in most (but still not
all) Russian cities were counted in a more fair way than in the parliamentary
elections. 

\begin{figure}[t]
\begin{centering}
\includegraphics[width=1\columnwidth]{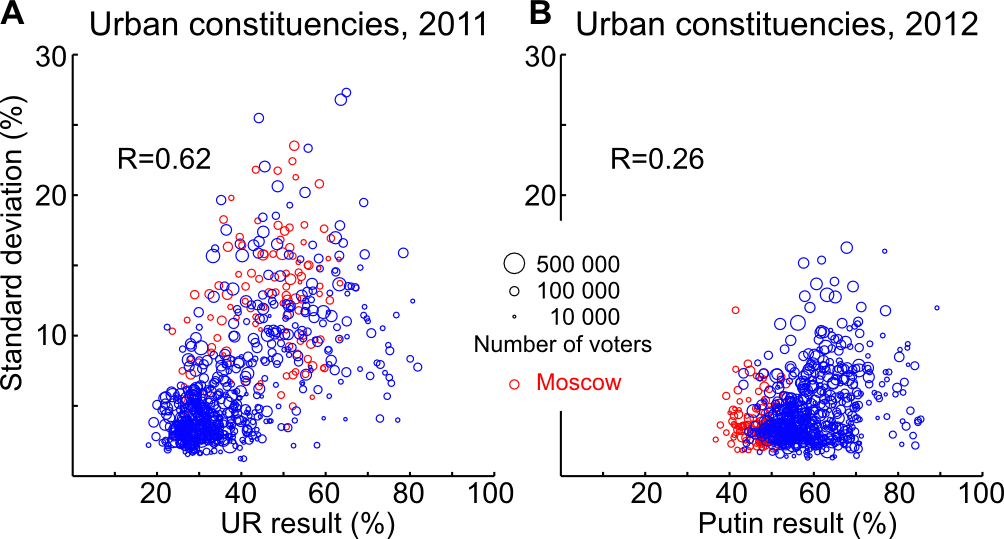}
\par\end{centering}

\caption{Standard deviations in 730 large urban constituencies (at least 8
polling stations with more than 1000 registered voters), excluding
the nine republics, for parliamentary (A) and presidential (B) elections.
Vertical axis shows the standard deviation of the results across polling
stations in a given constituency, while area of the circles is proportional
to the total number of registered voters in the constituency. $R$
stands for correlation coefficient. Note the sharp decrease of the
SD for Moscow constituencies (red circles) in presidential elections.
At the same time, in presidential elections 9 out of 10 constituencies
with the highest SD are located in the city of St. Petersburg.}

\end{figure}

To estimate the amount of ballots gained by the winners due to unusually
high correlation of their votes with turnout, we begin with the parliamentary
elections and consider the projection of the 2D diagram (Fig. 1A)
onto the horizontal axis (Fig. 1C). It looks similar to its vertical
counterpart (Fig. 1B), with sharp peaks at several round percentage
values and an extra maximum at large turnout. Note that, like in the
Moscow case (Fig. 2C), corresponding histograms for other parties
look quite different from that for UR, but very similar to each other.
The part of UR histogram that is not proportional to the cumulative
histogram of other parties (and is directly related to the positive
correlation of UR result with turnout) can easily be separated by
summing up votes for all parties except UR and rescaling the resulting
curve to fit the UR curve at lower turnouts, as shown schematically
in Fig. 1C. A more accurate calculation, performed individually for
urban and rural parts of every region (see Table S3 and Methods),
yields $\sim$11 million votes for UR (out of total 32.4 million)
associated with the turnout-UR correlation. One may speculate that
this part of ballots for UR was in some way {}``unfair'' (stuffed,
fraudulently counted, or obtained in non-voluntary voting settings).
If the applied procedure were entirely accurate, discarding these
votes would decrease the nationwide UR result to $\sim$39\%. However,
as some part of the observed correlation between UR result and turnout
could have arisen naturally (due to, for instance, social conformity
\cite{Coleman2004} or other confounding factors), this number probably
represents an upper estimate. The similar procedure applied to the
presidential elections yields a more modest result of $\sim$7 million
votes (out of total 45.6 million) for Putin, which is consistent with
the increased public control and official anti-forgery measures.

\begin{figure}[t]
\begin{centering}
\includegraphics[width=1\columnwidth]{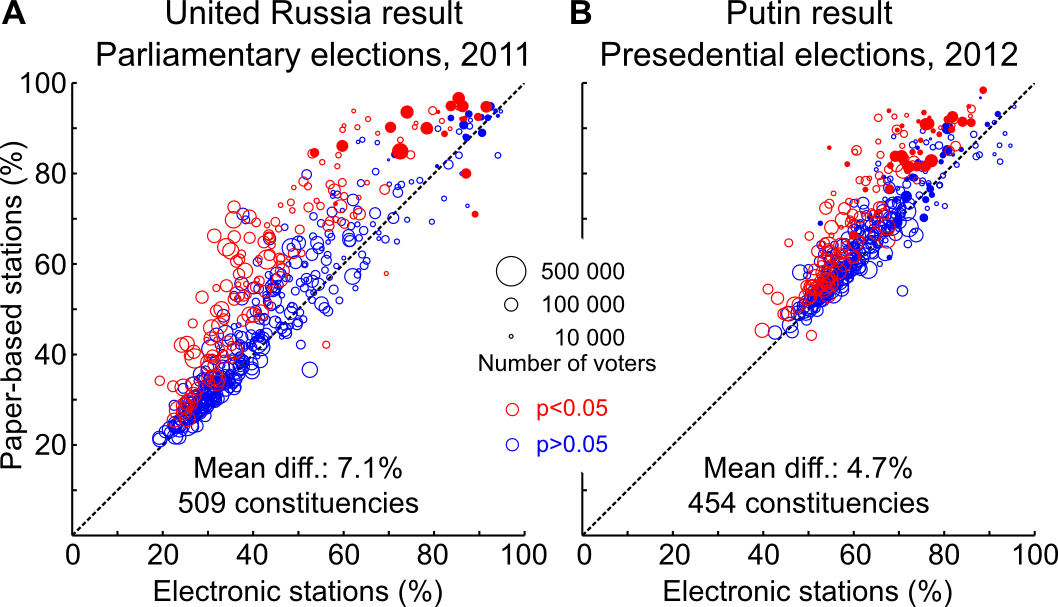}
\par\end{centering}

\caption{Correlation between winners' results at the electronic and the paper-based
polling stations at all constituencies with electronic polling stations
in parliamentary \textbf{(A)} and presidential \textbf{(B)} elections.
Circle areas are proportional to the number of registered voters in
a constituency. Filled circles show constituencies located in the
nine republics. Red circles show constituencies where UR/Putin results
at electronic and paper-based polling stations are significantly different
with $p<0.05$ (Mann-Whitney-Wilcoxon ranksum test); blue circles
show all the remaining constituencies.}
\end{figure}

Finally, at both elections, some polling stations ($\sim$5.5\% nationwide)
were equipped with electronic ballot boxes to scan the ballots and
count votes automatically, thereby reducing possibility of human interference.
Our analysis revealed (Fig. 4) that within the same constituencies
UR result at the electronic polling stations was on average 7.1\%
lower than at the traditional paper-based ones (difference significant
with $p=10^{-51}$, see Table S4 and Methods), and Putin\textquoteright{}s
result was 4.7\% lower ($p=10^{-35}$). While it cannot be taken for
granted that electronic polling stations constitute a representative
sub-ensemble, these differences are fairly consistent with our estimates
above.

Concluding, we have used the 2D correlation analysis to efficiently
pinpoint a number of anomalies in recent Russian elections, with a
short summary given by Table 1. Even though in all metrics discussed
the presidential elections appear to be fairer than the parliamentary
ones, various anomalies still amount to millions of ballots. While
statistical analysis per se does not (and cannot) serve as a concluding
proof of any possible fraud, it clearly highlights the alarming fingerprints
in the voting results.

\begin{table*}[t]
\begin{tabular*}{1\textwidth}{@{\extracolsep{\fill}}>{\centering}m{3cm}>{\centering}m{7.5cm}>{\centering}m{2cm}>{\centering}m{2cm}}
\multicolumn{2}{c}{\textbf{Parameter}} & \textbf{2011} & \textbf{2012}\tabularnewline[0.1cm]
\hline 
\noalign{\vskip0.1cm}
\multirow{3}{3cm}[-0.2cm]{\centering Result-turnout correlation for the pro-governmental candidate} & Computed over all polling stations & 0.68 & 0.53\tabularnewline[0.1cm]
\noalign{\vskip0.1cm}
 & Urban areas only & 0.44 & 0.29\tabularnewline[0.1cm]
\noalign{\vskip0.1cm}
 & Share of constituencies with significantly positive correlations ($p<0.05$) & 47\% & 35\%\tabularnewline[0.1cm]
\cline{2-4} 
\noalign{\vskip0.1cm}
\multirow{3}{3cm}{\centering Peaks at round numbers} & Area, millions of ballots & 1.4 & 1.3\tabularnewline[0.1cm]
\noalign{\vskip0.1cm}
 & Significance of the highest peak before 90\% & $p\approx10^{-19}$ & $p=5\cdot10^{-5}$\tabularnewline[0.1cm]
\noalign{\vskip0.1cm}
 & Joint significance of the peaks at 65\%... 85\% & $p\approx10^{-70}$ & $p=10^{-15}$\tabularnewline[0.1cm]
\cline{2-4} 
\noalign{\vskip0.1cm}
Anomalously low dispersion of results in a constituency & Number of constituencies with dispersion lower than the binomial one,
$p<0.0001$ & 25 & 9\tabularnewline[0.1cm]
\cline{2-4} 
\noalign{\vskip0.1cm}
Anomalously high dispersion of results in a constituency & Number of urban constituencies with standard deviation over 10\% & 185 & 28\tabularnewline[0.1cm]
\cline{2-4} 
\noalign{\vskip0.1cm}
\multirow{2}{3cm}[-0.3cm]{\centering Anomaly estimation} & Number of ballots & $\sim11\cdot10^{6}$ & $\sim7\cdot10^{6}$\tabularnewline[0.1cm]
\noalign{\vskip0.1cm}
 & Difference between percentage values of the official and estimated
results, percentage points  & $\sim10\%$ & $\sim4\%$\tabularnewline[0.1cm]
\cline{2-4} 
\noalign{\vskip0.1cm}
\multirow{2}{3cm}{\centering\emph{Koibatost}} & Averaged difference between the results at paper-based and electronic
polling stations & 7.1\% & 4.7\%\tabularnewline[0.1cm]
\noalign{\vskip0.1cm}
 & Significance of the difference & $p=10^{-51}$ & $p=10^{-35}$\tabularnewline[0.1cm]
\end{tabular*}

\caption{Anomalies in the voting data. The term \emph{koibatost} is derived
from a Russian name of the electronic ballot scanning device, KOIB.}
\end{table*}

\section*{Methods}

\subsection*{General background}

Seven parties participated in the parliamentary elections: United
Russia (49.3\%), Communist Party (19.2\%), A Just Russia (13.2\%),
Liberal Democratic Party (11.7\%), Yabloko (3.4\%), Patriots of Russia
(1.0\%), and Right Cause (0.6\%). Five candidates participated in
the presidential elections: Vladimir Putin (leader of United Russia,
63.6\%), Gennady Ziuganov (Communist Party, 17.1\%), Mikhail Prokhorov
(independent, 7.9\%), Vladimir Zhirinovsky (Liberal Democratic Party,
6.2\%), and Sergey Mironov (A Just Russia, 3.9\%).

\subsection*{Data acquisition}

The raw election data are officially available at Russian Central
Election Committee website (\emph{izbirkom.ru}) as multiple separate
HTML pages and Excel reports; the data from 95228/95416 (here and
below numbers refer to the parliamentary/presidential elections) polling
stations were downloaded programmatically to form a database. The
accuracy of the resulting databases was verified by checking regional
totals and comparing a number of randomly chosen polling stations
with the respective information at the official website. The list
of urban constituencies was composed by taking all 792 constituencies
conforming to certain name patterns (for instance, having the word
{}``city'' in the name) and manually adding 53 obviously urban constituencies
(total number of constituencies is 2744). Total number of ballots
cast in these urban constituencies was 37.1/41.1 million, and 28.3/30.1
million in the remaining (\textquotedblleft{}rural\textquotedblright{})
ones; additional 0.3/0.5 million ballots were collected abroad. Both
election databases along with the explanatory text are available in
the online supplementary materials. The nationwide lists of electronic
polling stations are not officially available. Therefore, the lists
of 4373/4943 polling stations with electronic ballot boxes in 72/76
regions of Russia were compiled of data gathered at the websites of
regional electoral committees (e.g. \emph{st-petersburg.izbirkom.ru/etc/138\_1pril.doc}
for St. Petersburg) and the government purchasing portal (e.g. \emph{zakupki.gov.ru/pgz/documentdownload?\linebreak
documentId=54880223} for Irkutsk region).

\subsection*{Data analysis}

To plot the curves presented in Figs. 1, 2, and S1, we added an artificial
white noise (uniformly distributed from \textminus{}0.5 to +0.5 votes)
to the number of ballots obtained by each party/candidate on each
polling station \cite{Johnston1995} and summed up the ballots within
a bin of 0.5\% for both turnout and result. The procedure was repeated
10 times, and the average was displayed. This eliminates possible
artefact peaks associated with division of integers (for example,
turnout is the ratio of two integer numbers).

\subsection*{Correlations}

In all cases, we use Spearman\textquoteright{}s correlation coefficients,
as they are more robust to outliers than the more conventional Pearson\textquoteright{}s
ones (e.g., military or hospital polling stations often behave like
outliers, with turnout close to 100\%; moreover, polling stations
located at the airports and train stations, where turnout is not defined,
are officially assigned the turnout of exactly 100\%). None of our
conclusions depend on this choice: we repeated all our analyses using
Pearson\textquoteright{}s correlation coefficients, and the difference
was always negligible (below 5\%).

\subsection*{Analysis of peaks}

The area under the peaks in Figs.1B,F was calculated as the area between
the actual curve and its smoothed version (filter cutoff frequency
0.2\%$^{-1}$, intervals \textpm{}2\% around each peak substituted
by a horizontal line segment before smoothing) in the intervals \textpm{}0.5\%
around each peak. The curve is quite noisy and so some peaks could
have appeared by chance; assuming this as a null hypothesis, we can
estimate the significance of peaks. First, standard deviation $\sigma$
was calculated as the root-mean-squared difference between the curve
and its smoothed version in the interval from 30\%/50\% to 90\% skipping
intervals \textpm{}2\% around peaks located at 65\%, 70\%, 75\%, 80\%
and 85\%, and the height of each peak $h_{i}$ was expressed in the
resulting $\sigma$ values. The $p$-values were then calculated as
$1-\textrm{erf}(\nicefrac{h_{i}}{\sqrt{2}})$, where erf denotes the
error function. For parliamentary elections the height of the 65\%
peak is $\sim9\sigma$, which corresponds to $p\approx10^{-19}$;
the product of $p$-values for the first 5 peaks we estimate to be
at least $10^{-70}$. For presidential elections the highest peak
is located at 75\% and is $3.9\sigma$ high ($p=5\cdot10^{-5}$);
the cumulative $p$-value for the same five peak positions is equal
to $10^{-15}$. As we are multiplying five separate $p$-values, the
values as low as $0.05{}^{5}\approx10^{-7}$ can still be considered
not significant; $p$-values obtained here are many orders of magnitude
lower than that.

\subsection*{Anomalously low variance of results per constituency}

First of all, we disregarded all polling stations with less than 50
registered voters (these are mostly temporary polling stations, often
located on ships, and therefore not representative of other polling
stations in the same constituency), and took all 2681 constituencies
with more than 5 remaining polling stations. For each of these constituencies,
we estimated the standard deviation of UR/Putin shares across polling
stations as median absolute deviation multiplied by 1.48 (median absolute
deviation is the median of deviations from the median; for a Gaussian
random variable it is 1.48 smaller than standard deviation) as a more
robust alternative to calculating standard deviation directly. If
$p$ is the median share and $n$ is the median number of ballots
across polling stations in the constituency, then the standard deviation
would be given by $\sqrt{p(1-p)/n}$, assuming the purely binomial
distribution of voting at every polling station with probability of
each person to vote for UR/Putin being $p$. As expected, in 97\%
of constituencies under consideration the observed standard deviation
was larger than the binomial one, which is the case if actual value
of $p$ varies across polling stations (for instance, due to local
inhomogeneities). However, in 83/87 constituencies the observed standard
deviation was smaller than the binomial one.

To estimate the statistical significance for each of these 83/87 constituencies,
we assume binomial voting as our null hypothesis, i.e. we assume that
on a polling station where the share of votes for UR/Putin is $p$,
every person votes for UR/Putin independently with probability $p$.
Let us now define $k$ as the number of polling stations in a constituency.
We take the half of the polling stations $\frac{k}{2}$ where the
UR/Putin share is closest to the median value of $p$, and set $p_{1}$
and $p_{2}$ as the minimal and maximal share in these $\frac{k}{2}$
polling stations. The probability $p_{0}$ to obtain a result between
$p_{1}$ and $p_{2}$ on a polling station with $n$ ballots, assuming
a binomial distribution of voting, can then be readily calculated
as $F(\lfloor np_{2}\rfloor,n,p)-F(\lfloor np_{1}\rfloor,n,p)$, where
$F$ is binomial cumulative distribution function (when $\lfloor np_{1}\rfloor$
was equal to $\lfloor np_{2}\rfloor$ we took $\lfloor np_{1}\rfloor-1$
instead). Finally, we calculate the $p$-value as the probability
to get at least $\frac{k}{2}$ successes out of $k$ trials with probability
of success being $p_{0}$, i.e. $F(\lfloor\frac{k}{2}\rfloor,k,p_{0})$.
There are 25/9 constituencies with $p<0.0001$ and 10/4 with $p<10^{-10}$.
Most notably, in 8/2 of these 25/9 constituencies the observed variance
is not only lower than the binomial one, but also the lowest possible:
at each of these $\frac{k}{2}$ polling stations the number of ballots
in favour of UR/Putin is given by multiplying the total number of
ballots by a fixed probability $p_{0}$, and rounding the result to
the nearest integer number (the resulting variance is non-zero only
because of this rounding). While theoretically this could have happened
by chance, in reality it is extremely unlikely. All of these 25/9
constituencies are located in the aforementioned nine republics (six
republics of North Caucasian Federal District, and Republics of Bashkortostan,
Tatarstan and Mordovia), which justifies considering them separately.

\subsection*{Standard deviations in urban constituencies}

The data presented in Fig. 3 are derived from all urban constituencies,
with the nine republics excluded. To calculate the standard deviation
in each constituency, we disregarded all polling stations with less
than 1000 registered voters. Smaller polling stations, that are not
typical for urban areas, are often situated in hospitals or military
zones, and therefore might substantially increase the standard deviation.
The 49 constituencies with less than eight remaining polling stations
were also omitted as it is not possible to reliably estimate standard
deviation with only few data points. This left 730 constituencies
to be analysed.

\subsection*{Estimating the amount of votes associated with the turnout-outcome
correlation}

Figure 1C shows the distribution f of votes in favour of United Russia
depending on the turnout, and the distribution g for the sum of votes
for all other parties. Until a threshold turnout of $\sim$50\%, these
two distributions are excellently proportional, $f=\alpha g$ (with
$\alpha$ being a scale coefficient), while at higher turnouts United
Russia\textquoteright{}s distribution starts to rise. The number of
additional UR ballots is thus given by $\sum(f-\alpha g)$. The computation
for the case of presidential elections is exactly the same.

We performed this analysis for every region, separately for urban
and rural parts, each time setting the turnout threshold in such a
way that 20\% of all ballots come from the polling stations with this
or lower turnout. This particular threshold value was chosen to reflect
the turnout intervals where the number of UR ballots is still proportional
to the sum of ballots for all other parties. Then $\alpha$ was found
with a least-squares fit, and the amount of additional UR/Putin ballots
was calculated by taking the sum starting from the threshold turnout.
Seven regions belonging to North Caucasian Federal District were analysed
altogether, with threshold turnout set manually to 75\%. In this Federal
District UR/Putin results at higher turnouts increase rapidly and
cease being proportional to the sum of votes for all other parties.

\subsection*{Analysis of results from electronic polling stations}

To calculate differences between UR/Putin results at paper-based and
electronic polling stations, we took all 509/454 (out of 2744) constituencies
that had at least two electronic and at least two paper-based stations.
In 422/371 of these constituencies, the joint UR/Putin result at all
traditional polling stations was higher than at all electronic ones
(see Fig. S3). The mode of the difference distribution was 0.2\%/0.7\%,
while the average difference was 7.1\%/4.7\%, which was significantly
higher than the mode with $p=10^{-51}$/$10^{-35}$ (Wilcoxon signed-rank
test). The slight non-zero mode of the distribution might be due to
some bias in how the electronic stations were located (e.g., in the
city centres, where UR/Putin support might have been lower than in
the city outskirts).

\section*{Acknowledgements}

We thank S.Slyusarev and B.Ovchinnikov for comments and suggestions,
and A.Shipilev for providing preliminary election data on the fly.
B.Ovchinnikov is especially acknowledged for drawing our attention
to high dispersion across polling stations in each constituency as
one of the metrics for election anomalies.

\bibliographystyle{unsrt}
\bibliography{elect}

\begin{thebibliography}{1}

\bibitem{Mikhailov2004}
V.~Mikhailov.
\newblock Regional elections and democratization in russia.
\newblock In Cameron Ross, editor, {\em Russian Politics under Putin}.
  Manchester University Press, 2004.

\bibitem{Myagkov2009}
M.G. Myagkov, P.C. Ordershook, and D.~Shakin.
\newblock {\em The forensics of Election Fraud: Russia and Ukraine}.
\newblock Cambridge University Press, 2009.

\bibitem{Klimek2012}
P.~Klimek, Y.~Yegorov, R.~Hanel, and S.~Thurner.
\newblock It's not the voting that's democracy, it's the counting: Statistical
  detection of systematic election irregularities.
\newblock {\em Arxiv preprint arXiv:1201.3087}, 2012.

\bibitem{Hamm2011}
P.~Hamm and M.~Zanni.
\newblock {\em Concepts and Methods of 2D Infrared Spectroscopy}.
\newblock Cambridge University Press, 2011.

\bibitem{Roy2011}
S.~Roy, M.~S. Pshenichnikov, and T.~L.~C. Jansen.
\newblock Analysis of 2d cs spectra for systems with non-gaussian dynamics.
\newblock {\em J. Phys. Chem. B.}, 115:5431, 2011.

\bibitem{Hansford2010}
T.G. Hansford and B.T. Gomez.
\newblock Estimating the electoral effects of voter turnout.
\newblock {\em American Political Science Review}, 104:268--288, 2010.

\bibitem{Coleman2004}
S.~Coleman.
\newblock The effect of social conformity on collective voting.
\newblock {\em Polit. Anal.}, 12:76--96, 2004.

\bibitem{Johnston1995}
R.G. Johnston, S.D. Schroder, and A.R. Mallawaaratchy.
\newblock Statistical artifacts in the ratio of discrete quantities.
\newblock {\em The American Statistician}, 49:285--291, 1995.

\end{thebibliography}
\newpage{}

\section*{Supplementary Discussion}

\subsection*{1. Correlation strength at different aggregation levels}

Even though there is strong positive correlation between turnout and
UR/Putin result in the nationwide data, in some regions this correlation
is absent or even negative. In 11/20 regions (here and below: parliamentary/presidential
elections) the urban part demonstrates significant ($p<0.05$) negative
correlation between turnout and UR/Putin result, and in 27/ 23 regions
urban correlation does not significantly differ from zero ($p>0.05$).
For rural parts these numbers are 2/1 and 4/6, respectively. In general,
correlation increases at higher aggregation levels: if all individual
polling stations are considered, the correlation coefficient is 0.68/0.53,
as stated in the main text; taking all constituencies as data points
yields the result of 0.80/0.63; taking all regions --- 0.82/0.69. 

Intraregional correlations between turnout and UR/Putin results do
not arise due to aggregation of different constituencies: these correlations
can already be observed inside individual constituencies. To show
that, for both urban and rural parts of every region we computed the
overall correlation coefficient $R_{i}$ and the constituency-level
correlation coefficient $Q_{i}$, given by computing correlation coefficients
inside each constituency and averaging over constituencies. Values
of $R_{i}$ and $Q_{i}$ were highly correlated with correlation coefficient
of 0.85/0.89 and regression slope of 0.74/0.77. Overall, positive
and significant ($p<0.05$) correlation is present inside 46\%/35\%
of all constituencies (in 23\%/15\% for $p<0.001$), as opposed to
only 3\%/4\% showing significant ($p<0.05$) negative correlations.

\subsection*{2. Relation between high \emph{koibatost} and high standard deviation
on the constituency level}

For urban constituencies in the parliamentary elections there is a
high correlation (0.66) between standard deviation of UR results and
paper-electronic difference (calculated over 264 constituencies where
the data are available, see Methods). Moreover, the same constituency
(in the city of Magnitogorsk) holds the top positions according to
both criteria (see Tables S2 and S4, which can hardly be a coincidence.
This additionally proves that high standard deviation is indeed a
useful metric for election anomalies.

\subsection*{3. Urban-rural separation}

One point of concern with the urban-rural separation is that only
the polling stations from fully urban constituencies are classified
as urban. As a result, \textquotedblleft{}rural\textquotedblright{}
part still contains numerous small towns. This might induce a spurious
correlation between turnout and UR/Putin results, as smaller settlements
tend to demonstrate higher turnout and higher UR/Putin results. To
address this issue, we separated \textquotedblleft{}rural\textquotedblright{}
part of each region into two parts: large rural polling stations with
the number of registered voters over 950 (mostly small towns and large
villages), and small rural polling stations with the number of registered
voters less than 950 (mostly small villages). The 950 threshold was
chosen because the distribution of polling stations by the number
of registered voters is bimodal with a node around 950. Such an approach
indeed reduces \textquotedblleft{}rural\textquotedblright{} turnout-UR
correlations (for instance, for the parliamentary elections from 0.64
to 0.58), but the overall estimate of the number of votes associated
with correlations, when computed separately for urban, large rural
and small rural polling stations in each region, remains almost the
same (for the parliamentary elections the number slightly decreased
from 11 million to 10.5 million).

\section*{Supplementary Figures and Tables}

See next page.

\setcounter{figure}{0}
\makeatletter
\renewcommand{\thefigure}{S\@arabic\c@figure}  
\makeatother 

\setcounter{table}{0}
\makeatletter
\renewcommand{\thetable}{S\@arabic\c@table}  
\makeatother 

\begin{table*}
\begin{tabular*}{1\textwidth}{@{\extracolsep{\fill}}>{\centering}m{0.7cm}>{\centering}m{5cm}>{\centering}m{1.5cm}>{\centering}m{5cm}>{\centering}m{1.5cm}}
\noalign{\vskip0.1cm}
 & \multicolumn{2}{c}{\textbf{\footnotesize 2011}} & \multicolumn{2}{c}{\textbf{\footnotesize 2012}}\tabularnewline[0.1cm]
\noalign{\vskip0.1cm}
 & \textbf{\footnotesize Region and constituency} & \textbf{\emph{\footnotesize p}} & \textbf{\footnotesize Region and constituency} & \textbf{\emph{\footnotesize p}}\tabularnewline[0.1cm]
\cline{2-5} 
\noalign{\vskip0.1cm}
{\footnotesize 1} & {\footnotesize Respublika Dagestan, Dahadaevskaja } & {\footnotesize $<10^{-15}$} & {\footnotesize Respublika Severnaja Osetija, Levoberezhnoj chasti
g.Vladikavkaza } & {\footnotesize $2\cdot10^{-14}$}\tabularnewline[0.1cm]
\noalign{\vskip0.1cm}
{\footnotesize 2} & {\footnotesize Kabardino-Balkarskaja Respublika, Prohladnenskaja } & {\footnotesize $1\cdot10^{-15}$} & {\footnotesize Respublika Dagestan, Derbentskaja gorodskaja } & {\footnotesize $1\cdot10^{-13}$}\tabularnewline[0.1cm]
\noalign{\vskip0.1cm}
{\footnotesize 3} & {\footnotesize Respublika Dagestan, Sulejman-Stal'skaja } & {\footnotesize $5\cdot10^{-15}$} & {\footnotesize Respublika Dagestan, Kiziljurtovskaja} & {\footnotesize $1\cdot10^{-12}$}\tabularnewline[0.1cm]
\noalign{\vskip0.1cm}
{\footnotesize 4} & {\footnotesize Respublika Dagestan, Mahachkala, Sovetskaja } & {\footnotesize $4\cdot10^{-14}$} & {\footnotesize Kabardino-Balkarskaja Respublika, Prohladnenskaja} & {\footnotesize $7\cdot10^{-11}$}\tabularnewline[0.1cm]
\noalign{\vskip0.1cm}
{\footnotesize 5} & {\footnotesize Respublika Dagestan, Babajurtovskaja } & {\footnotesize $4\cdot10^{-13}$} & {\footnotesize Respublika Dagestan, Hunzahskaja } & {\footnotesize $2\cdot10^{-10}$}\tabularnewline[0.1cm]
\noalign{\vskip0.1cm}
{\footnotesize 6} & {\footnotesize Respublika Bashkortostan, Sterlitamakskaja gorodskaja} & {\footnotesize $6\cdot10^{-13}$} & {\footnotesize Respublika Dagestan, Kizljarskaja} & {\footnotesize $2\cdot10^{-9}$}\tabularnewline[0.1cm]
\noalign{\vskip0.1cm}
{\footnotesize 7} & {\footnotesize Respublika Severnaja Osetija, Levoberezhnoj chasti
g.Vladikavkaza } & {\footnotesize $2\cdot10^{-12}$} & {\footnotesize Respublika Tatarstan, Zainskaja} & {\footnotesize $6\cdot10^{-6}$}\tabularnewline[0.1cm]
\noalign{\vskip0.1cm}
{\footnotesize 8} & {\footnotesize Respublika Dagestan, Sergokalinskaja} & {\footnotesize $2\cdot10^{-12}$} & {\footnotesize Kabardino-Balkarskaja Respublika, Baksanskaja} & {\footnotesize $3\cdot10^{-5}$}\tabularnewline[0.1cm]
\noalign{\vskip0.1cm}
{\footnotesize 9} & {\footnotesize Respublika Dagestan, Hunzahskaja} & {\footnotesize $6\cdot10^{-12}$} & {\footnotesize Respublika Tatarstan, Nurlatskaja} & {\footnotesize $6\cdot10^{-5}$}\tabularnewline[0.1cm]
\noalign{\vskip0.1cm}
{\footnotesize 10} & {\footnotesize Respublika Dagestan, Kizljarskaja } & {\footnotesize $8\cdot10^{-12}$} & {\footnotesize Respublika Dagestan, Bezhtinskaja } & {\footnotesize $3\cdot10^{-4}$}\tabularnewline[0.1cm]
\end{tabular*}

\caption{Top ten constituencies with the most anomalously low dispersions}
\end{table*}

\begin{table*}
\begin{tabular*}{1\textwidth}{@{\extracolsep{\fill}}>{\centering}m{0.7cm}>{\centering}m{5cm}>{\centering}m{1.5cm}>{\centering}m{5cm}>{\centering}m{1.5cm}}
\noalign{\vskip0.1cm}
 & \multicolumn{2}{c}{\textbf{\footnotesize 2011}} & \multicolumn{2}{c}{\textbf{\footnotesize 2012}}\tabularnewline[0.1cm]
\noalign{\vskip0.1cm}
 & \textbf{\footnotesize Region and constituency} & \textbf{\emph{\footnotesize p}} & \textbf{\footnotesize Region and constituency} & \textbf{\emph{\footnotesize p}}\tabularnewline[0.1cm]
\cline{2-5} 
\noalign{\vskip0.1cm}
{\footnotesize 1} & {\footnotesize Cheljabinskaja oblast', Magnitogorsk, Pravoberezhnaja} & {\footnotesize 27.3\%} & {\footnotesize St. Petersburg, \#17} & {\footnotesize 16.3\%}\tabularnewline[0.1cm]
\noalign{\vskip0.1cm}
{\footnotesize 2} & {\footnotesize Cheljabinskaja oblast', Magnitogorsk, Ordzhonikidzevskaja} & {\footnotesize 26.8\%} & {\footnotesize Krasnodarskij kraj, Novorossijsk, Vostochnaja} & {\footnotesize 16.0\%}\tabularnewline[0.1cm]
\noalign{\vskip0.1cm}
{\footnotesize 3} & {\footnotesize Vladimirskaja oblast', Vladimir, Oktjabr'skaja} & {\footnotesize 25.5\%} & {\footnotesize St. Petersburg, \#30} & {\footnotesize 15.4\%}\tabularnewline[0.1cm]
\noalign{\vskip0.1cm}
{\footnotesize 4} & {\footnotesize Moscow, rajon Gol'janovo} & {\footnotesize 23.5\%} & {\footnotesize St. Petersburg, \#19} & {\footnotesize 15.2\%}\tabularnewline[0.1cm]
\noalign{\vskip0.1cm}
{\footnotesize 5} & {\footnotesize Cheljabinskaja oblast', Magnitogorsk, Leninskaja} & {\footnotesize 23.3\%} & {\footnotesize St. Petersburg, \#27} & {\footnotesize 13.9\%}\tabularnewline[0.1cm]
\noalign{\vskip0.1cm}
{\footnotesize 6} & {\footnotesize Moscow, rajon Severnoe Butovo} & {\footnotesize 22.4\%} & {\footnotesize St. Petersburg, \#2} & {\footnotesize 13.7\%}\tabularnewline[0.1cm]
\noalign{\vskip0.1cm}
{\footnotesize 7} & {\footnotesize Vladimirskaja oblast', Kovrovskaja gorodskaja} & {\footnotesize 22.0\%} & {\footnotesize St. Petersburg, \#1} & {\footnotesize 13.5\%}\tabularnewline[0.1cm]
\noalign{\vskip0.1cm}
{\footnotesize 8} & {\footnotesize Moscow, rajon Hamovniki} & {\footnotesize 21.8\%} & {\footnotesize St. Petersburg, \#11} & {\footnotesize 12.9\%}\tabularnewline[0.1cm]
\noalign{\vskip0.1cm}
{\footnotesize 9} & {\footnotesize Moscow, rajon Bogorodskoe} & {\footnotesize 21.7\%} & {\footnotesize St. Petersburg, \#24} & {\footnotesize 12.8\%}\tabularnewline[0.1cm]
\noalign{\vskip0.1cm}
{\footnotesize 10} & {\footnotesize Moscow, rajon Prospekt Vernadskogo} & {\footnotesize 21.3\%} & {\footnotesize St. Petersburg, \#29} & {\footnotesize 12.8\%}\tabularnewline[0.1cm]
\end{tabular*}

\caption{Top ten urban constituencies with largest standard deviations (SDs) }
\end{table*}

\begin{table*}
\begin{tabular*}{1\textwidth}{@{\extracolsep{\fill}}>{\centering}m{0.7cm}>{\centering}m{5cm}>{\centering}m{1.5cm}>{\centering}m{5cm}>{\centering}m{1.5cm}}
\noalign{\vskip0.1cm}
 & \multicolumn{2}{c}{\textbf{\footnotesize 2011}} & \multicolumn{2}{c}{\textbf{\footnotesize 2012}}\tabularnewline[0.1cm]
\noalign{\vskip0.1cm}
 & \textbf{\footnotesize Region} & \textbf{\footnotesize Correlation- related votes} & \textbf{\footnotesize Region} & \textbf{\footnotesize Correlation- related votes}\tabularnewline[0.1cm]
\cline{2-5} 
\noalign{\vskip0.1cm}
{\footnotesize 1} & {\footnotesize Six republics of North Caucasus} & {\footnotesize 2 300 000} & {\footnotesize Six republics of North Caucasus} & {\footnotesize 1 800 000}\tabularnewline[0.1cm]
\noalign{\vskip0.1cm}
{\footnotesize 2} & {\footnotesize Moscow} & {\footnotesize 1 000 000} & {\footnotesize Respublika Tatarstan} & {\footnotesize 650 000}\tabularnewline[0.1cm]
\noalign{\vskip0.1cm}
{\footnotesize 3} & {\footnotesize Respublika Bashkortostan} & {\footnotesize 790 000} & {\footnotesize Respublika Bashkortostan} & {\footnotesize 570 000}\tabularnewline[0.1cm]
\noalign{\vskip0.1cm}
{\footnotesize 4} & {\footnotesize Respublika Tatarstan} & {\footnotesize 770 000} & {\footnotesize Kemerovskaja oblast'} & {\footnotesize 440 000}\tabularnewline[0.1cm]
\noalign{\vskip0.1cm}
{\footnotesize 5} & {\footnotesize Krasnodarskij kraj} & {\footnotesize 580 000} & {\footnotesize Krasnodarskij kraj} & {\footnotesize 410 000}\tabularnewline[0.1cm]
\noalign{\vskip0.1cm}
{\footnotesize 6} & {\footnotesize Saratovskaja oblast'} & {\footnotesize 450 000} & {\footnotesize Nizhegorodskaja oblast'} & {\footnotesize 280 000}\tabularnewline[0.1cm]
\noalign{\vskip0.1cm}
{\footnotesize 7} & {\footnotesize Kemerovskaja oblast'} & {\footnotesize 410 000} & {\footnotesize St. Petersburg } & {\footnotesize 260 000}\tabularnewline[0.1cm]
\noalign{\vskip0.1cm}
{\footnotesize 8} & {\footnotesize Respublika Mordovija} & {\footnotesize 360 000} & {\footnotesize Saratovskaja oblast'} & {\footnotesize 240 000}\tabularnewline[0.1cm]
\noalign{\vskip0.1cm}
{\footnotesize 9} & {\footnotesize Rostovskaja oblast'} & {\footnotesize 320 000} & {\footnotesize Respublika Mordovija} & {\footnotesize 210 000}\tabularnewline[0.1cm]
\noalign{\vskip0.1cm}
{\footnotesize 10} & {\footnotesize Voronezhskaja oblast'} & {\footnotesize 260 000} & {\footnotesize Primorskij kraj} & {\footnotesize 160 000}\tabularnewline[0.1cm]
\end{tabular*}

\caption{Top ten regions with largest amounts of correlation-related votes}
\end{table*}

\begin{table*}
\begin{tabular*}{1\textwidth}{@{\extracolsep{\fill}}>{\centering}m{0.7cm}>{\centering}m{5cm}>{\centering}m{1.5cm}>{\centering}m{5cm}>{\centering}m{1.5cm}}
\noalign{\vskip0.1cm}
 & \multicolumn{2}{c}{\textbf{\footnotesize 2011}} & \multicolumn{2}{c}{\textbf{\footnotesize 2012}}\tabularnewline[0.1cm]
\noalign{\vskip0.1cm}
 & \textbf{\footnotesize Region and constituency} & \textbf{\emph{\footnotesize Koibatost}} & \textbf{\footnotesize Region and constituency} & \textbf{\emph{\footnotesize Koibatost}}\tabularnewline[0.1cm]
\cline{2-5} 
\noalign{\vskip0.1cm}
{\footnotesize 1} & {\footnotesize Cheljabinskaja oblast, Magnitogorsk, Pravoberezhnaja} & {\footnotesize 36.8\%} & {\footnotesize Respublika Bashkortostan, Kiginskaja} & {\footnotesize 31.1\%}\tabularnewline[0.1cm]
\noalign{\vskip0.1cm}
{\footnotesize 2} & {\footnotesize Astrahanskaja oblast, Astrahan, Leninskaja} & {\footnotesize 34.8\%} & {\footnotesize Astrahanskaja oblast', Privolzhskaja} & {\footnotesize 26.9\%}\tabularnewline[0.1cm]
\noalign{\vskip0.1cm}
{\footnotesize 3} & {\footnotesize Cheljabinskaja oblast', Magnitogorsk, Ordzhonikidzevskaja} & {\footnotesize 34.1\%} & {\footnotesize Respublika Bashkortostan, Belokatajskaja} & {\footnotesize 26.1\%}\tabularnewline[0.1cm]
\noalign{\vskip0.1cm}
{\footnotesize 4} & {\footnotesize Astrahanskaja oblast', Astrahan', Kirovskaja } & {\footnotesize 33.3\%} & {\footnotesize Tjumenskaja oblast', Kazanskaja} & {\footnotesize 24.5\%}\tabularnewline[0.1cm]
\noalign{\vskip0.1cm}
{\footnotesize 5} & {\footnotesize Tjumenskaja oblast', Jurginskaja} & {\footnotesize 31.6\%} & {\footnotesize Voronezhskaja oblast', Cemilukskaja} & {\footnotesize 24.5\%}\tabularnewline[0.1cm]
\noalign{\vskip0.1cm}
{\footnotesize 6} & {\footnotesize Saratovskaja oblast', Petrovskaja} & {\footnotesize 31.4\%} & {\footnotesize Tjumenskaja oblast', Abatskaja} & {\footnotesize 23.6\%}\tabularnewline[0.1cm]
\noalign{\vskip0.1cm}
{\footnotesize 7} & {\footnotesize Saratovskaja oblast', Rtiwevskaja} & {\footnotesize 31.3\%} & {\footnotesize Respublika Bashkortostan, Kugarchinskaja} & {\footnotesize 23.5\%}\tabularnewline[0.1cm]
\noalign{\vskip0.1cm}
{\footnotesize 8} & {\footnotesize Tjumenskaja oblast', Tjumen', Vostochnaja} & {\footnotesize 31.3\%} & {\footnotesize Tjumenskaja oblast', Omutinskaja} & {\footnotesize 21.7\%}\tabularnewline[0.1cm]
\noalign{\vskip0.1cm}
{\footnotesize 9} & {\footnotesize Respublika Mordovija, Ruzaevskaja} & {\footnotesize 31.0\%} & {\footnotesize Tjumenskaja oblast', Jurginskaja} & {\footnotesize 21.2\%}\tabularnewline[0.1cm]
\noalign{\vskip0.1cm}
{\footnotesize 10} & {\footnotesize Tjumenskaja oblast', Sorokinskaja} & {\footnotesize 30.5\%} & {\footnotesize Saratovskaja oblast', Marksovskaja} & {\footnotesize 20.2\%}\tabularnewline[0.1cm]
\end{tabular*}

\caption{Top ten constituencies with largest values of \emph{koibatost}. \emph{Koibatost}
refers to the difference between the results at paper-based and electronic
polling stations.}
\end{table*}

\begin{figure*}
\begin{centering}
\includegraphics[width=1\textwidth]{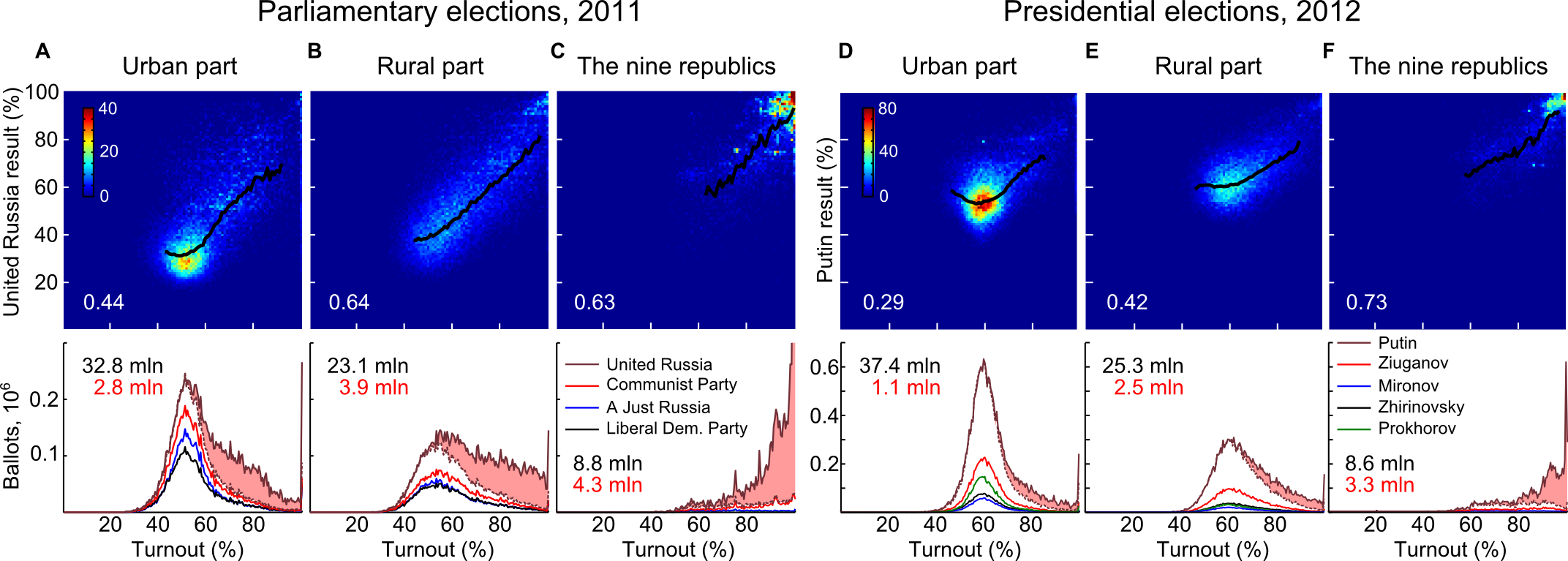}
\par\end{centering}

\caption{Decomposition of two-dimensional histogram of UR \textbf{(A--C)} and
Putin \textbf{(D--F)} votes shown in Figs. 1A,E into three parts:
urban territories \textbf{(A,D)}, rural territories \textbf{(B,E)},
and the nine republics \textbf{(C,F)} that form a separate cluster
at very high turnout values (see text). Black lines show overall result
for each turnout bin; white numbers stand for correlation coefficients.
Horizontal projections in the lower panel are analogous to Fig. 1C
and show total number votes depending on the turnout (0.5\% bin).
Black numbers represent total number of ballots in these areas, red
numbers show the amount of votes associated with turnout-result correlation.
Red shading is only an illustrative sketch as the actual calculations
were performed for each region separately (see text). The colour code
corresponds to thousands of votes in a $1\times1\%$ bin. Note the
shining dot in (D) at 60\% turnout and 80\% result that can be traced
to the city of St. Petersburg and comprises $\sim$36.5 thousand votes
for Putin (2.6\% of the city total votes).}
\end{figure*}

\end{document}